# Fermi Condensates

Markus Greiner, Cindy A. Regal, and Deborah S. Jin

*JILA, National Institute of Standards and Technology and University of Colorado, and Department of Physics, University of Colorado, Boulder, CO 80309-0440*

**Abstract.** Ultracold atomic gases have proven to be remarkable model systems for exploring quantum mechanical phenomena. Experimental work on gases of fermionic atoms in particular has seen large recent progress including the attainment of so-called Fermi condensates. In this article we will discuss this recent development and the unique control over interparticle interactions that made it possible.

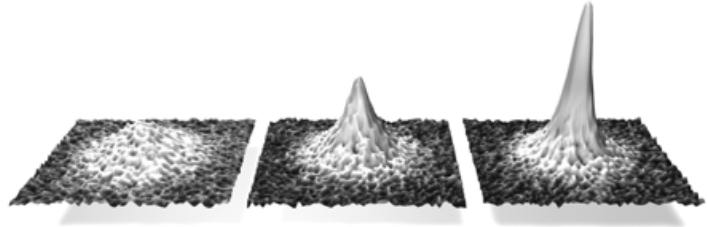

**FIGURE 1:** Fermionic condensates, shown in surface plots of time-of-flight absorption images. These condensates of generalized Cooper pairs are created at different interaction strengths in the strongly interacting BCS-BEC crossover regime [1].

The fascinating aspect about experiments with ultracold atoms is that it has become possible to create complex, but yet very accessible and well controlled many-body quantum systems. Bose-Einstein condensates (BEC) [2, 3] have been used to study effects such as coherence, vortices or superfluid flow, and a quantum phase transition from a superfluid to a Mott insulator has been observed [4]. Recently, experiments have come to a point where it becomes possible to also study ultracold Fermionic atoms. After creating a degenerate Fermi sea in our group[5], work has concentrated on manipulating the interatomic interaction using magnetic Feshbach resonances [6-11]. In 2003, molecular Bose-Einstein condensates have been created [12-14], where the weakly bound molecules are formed of two fermionic atoms. These experiments were soon followed by the observation of Fermionic condensates of generalized Cooper Pairs in the BCS-BEC crossover regime (Fig. 1) [1, 15].

## ULTRACOLD QUANTUM GASES

A system behaves quantum mechanically when the matter wavelength of the particles is of comparable size to the typical spacing between particles. For a gas of atoms, with relatively heavy particles and low density, exploring quantum behavior requires that we cool the atom gas to extraordinarily low temperatures near absolute zero. This is typically achieved with a combination of laser cooling and trapping followed by evaporative cooling in either a magnetic or optical trap. A typical experiment can produce a million atoms at temperatures around 100 nK and densities between $10^{13}$ and $10^{14}$ atoms/cm$^3$. Only at these very low temperatures does it become important that the atoms making up our gas are either bosons or fermions, which are the two classes of quantum particles found in nature. Bosons have integer spin and prefer to occupy identical quantum states. If they are trapped at low enough temperatures, they macroscopically occupy the lowest possible energy state and form a Bose-Einstein condensate (BEC). This condensation elevates quantum behavior to a macroscopic scale and results in fascinating phenomena such as coherent matter waves and superfluidity.

Fermions, on the other hand, have half integer spin and constitute the second half of the particle family tree. To tell whether an atom is a boson or a fermion, we can simply look at the total number of protons, neutrons, and electrons making up that atom. Since these are all spin ½ fermions, adding up an odd number of them will make an atom that is a fermion (half integer spin). Conversely, an even total number will make an atom that is a boson (integer spin). Fermions obey the Pauli exclusion principle. This means that two indistinguishable Fermions can never occupy the same quantum state. Instead, in the limit of absolute zero temperature they fill the lowest energy states of the trap with exactly one atom per state in an arrangement known as the Fermi sea (Fig. 2). In the particular experiments we will discuss here, the gas consists of a roughly equal mixture of two different spin states (internal states of the atom) so there are two atoms per energy state of the trap, with one in each spin state.

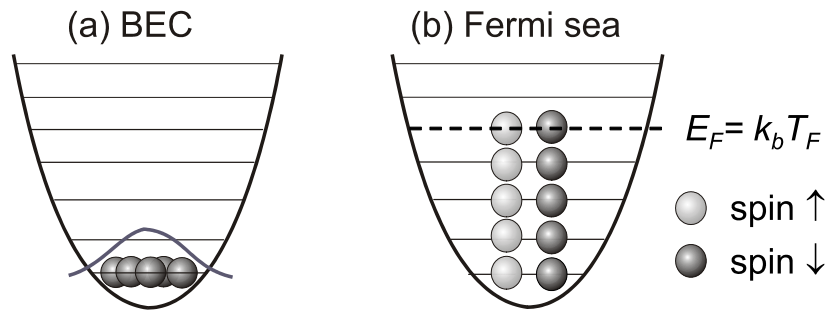

**FIGURE 2:** Bosonic and fermionic particles, e.g. ultracold atoms, in a confining harmonic potential. (a) Bosonic atoms macroscopically occupy the lowest possible energy state and form a Bose-Einstein condensate. They are superfluid and can be collectively described by a single macroscopic wave function. (b) Fermionic particles obey the Pauli exclusion principle and form a Fermi sea at temperatures close to absolute zero. All energy states up to the Fermi energy $E_F$ are filled, here with particles in two different spin states. A Fermi sea is not a superfluid.

Experimentally, these effects are seen by taking optical images of the gas. By suddenly turning off the trap and allowing the gas to expand for 10 to 20 ms before taking a snapshot, we measure the velocity distribution of the originally trapped gas. A Bose-Einstein condensate shows up dramatically in such an image as a large spike near the center of the expanded cloud corresponding to many atoms at near zero velocity. A Fermi sea of atoms shows up as a more subtle distortion of the classical gaussian velocity distribution but can be clearly detected in the measured velocity distributions. Since the imaging destroys the ultracold gas, data is accumulated by repeating the entire cooling and imaging procedure many times.

## BOSE-EINSTEIN CONDENSATION AND FERMIONS

The general quantum mechanical phenomenon of Bose-Einstein condensation shows up in many areas of physics, ranging from condensed matter physics, nuclear physics, elementary particle physics, astrophysics, and atomic physics. Examples of this phenomenon cover a tremendously range of size and temperature scales. Some examples are Cooper pairs of electrons in superconductors, $^4$He atoms in superfluid liquid He, excitons or biexcitons in semiconductors, neutron pairs or proton pairs in nuclei and also in neutron stars, mesons in neutron star matter, $^3$He atom pairs in superfluid $^3$He, as well as alkali atoms in ultracold atom gases.

Bose-Einstein condensation is a behavior of bosons and not of fermions. However, we see that, more often than not, the boson that condenses in the above examples is actually a pair of fermions, such as Cooper pairs of electrons or pairs of protons or neutrons in nuclei. This is not surprising since visible matter is made up of spin ½ fermions, and a pair of fermions is the smallest even number of fermions one could use to create a composite boson. An interesting question is then how this bosonic degree of freedom emerges from the underlying fermionic degrees of freedom. In particular, under certain conditions the fermionic nature of the constituent particles and the bosonic character of pairs can both play essential roles in the condensation phenomenon.

If we start with a gas of bosonic atoms, such as $^{87}$Rb or $^7$Li, we can only explore the behavior of bosons. The underlying fermion degrees of freedom are completely irrelevant since the energy cost to break the atom into two fermions (an electron and an ion) is 10 orders of magnitude larger than the condensation energy. If instead we start with a gas of fermionic atoms, such as $^{40}$K or $^6$Li, it turns out that we can experimentally explore the connection between fermionic superfluidity, such as superconductivity, and Bose-Einstein condensation by directly controlling the interactions between atoms and creating conditions that favor pairing.

## CONTROLLING INTERACTIONS

A key recent advance in quantum gas experiments is the ability to control the interactions between atoms. This unique control lies at the heart of recent experiments that make condensates starting with a Fermi gas of atoms. Not only is tuning the interaction strength essential for creating these condensates, it is moreover a unique and powerful tool for experimentally investigating the intriguing connection between superconductivity and Bose-Einstein condensation.

The handle with which we can control the interactions between atoms is called a magnetic-field Feshbach resonance. Around special values of a magnetic field, relatively small changes in magnetic field strength can have dramatic effects on the effective interactions in an ultracold gas. On one side of the Feshbach resonance value, for example a slightly higher magnetic field strength in the case of our experiments with $^{40}$K atoms, the gas has very strong, effectively attractive interactions between the atoms. It was predicted fairly recently that these attractive interactions would result in condensation of fermionic atom pairs and superfluidity of gas at unexpectedly high (but still ultracold) temperatures [16].

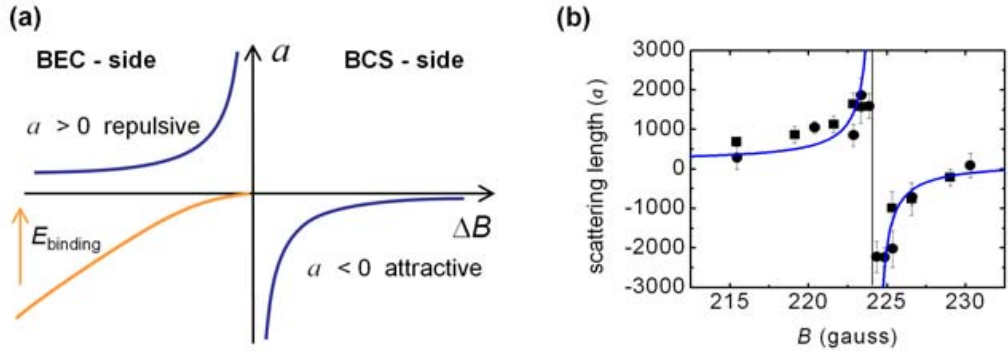

**FIGURE 3:** The interatomic interaction can be widely varied with a magnetic Feshbach resonance. (a) At a Feshbach resonance, a new molecular bound state is formed as the *B*-field is ramped across the resonance. The binding energy of this weakly bound state is indicated by the lower left line. The emergence of a new bound state leads to the divergence of the interaction properties. The interaction can be quantified with the *s*-wave scattering length *a*. Above the resonance, on the BCS-side, the interaction is attractive (*a*<0). Below the resonance (BEC-side), where the new molecular state is formed, the effective interatomic interaction is repulsive (*a*>0). Close to resonance the interaction is very strong and *a* diverges. In this regime, the molecular binding energy $E_b$ is fundamentally linked to the scattering length: $E_b = \hbar^2/(ma^2)$, where *m* is the atom mass. This single-channel picture of a Feshbach resonance describes the physics for a broad Feshbach resonance, as the ones used in present experiments. (b) We have measured the scattering length versus the *B*-field by measuring a mean-field shift with rf spectroscopy and observed the characteristic divergence of the scattering length [7].

To further elucidate the Feshbach resonance physics let us consider two atoms in vacuum. Exactly at the magnetic-field value of the Feshbach resonance, a new molecule state appears with an energy exactly equal to the energy of two free atoms (Fig. 3). As the magnetic-field strength is tuned away from the resonance on one side (lower values in our experiments), the energy of this new molecule state decreases (increasing binding energy). It should be emphasized that this is a very exotic diatomic molecule that is large and extremely weakly bound. The binding energy is on the order of tens of kHz. Furthermore the binding energy, and therefore size, of this molecule can be tuned simply by varying the magnetic-field strength a small amount near the Feshbach resonance.

In our experiments, as in a previous experiment using bosonic atoms [17], we were able to create large numbers of ultracold molecules using time-dependent magnetic-field sweeps around a Feshbach resonance [9, 10]. In our case of fermionic atoms, the resonant interactions and therefore the molecules involve atoms in two different spin states. We measured the binding energy of these molecules using rf photodissociation. (Note that dissociation of the molecules with an rf photon rather than the usual optical photon is only possible because of their ridiculously small binding energies.) As predicted we observed that the binding energy depends on magnetic-field strength near the Feshbach resonance (Fig. 4). Further, our measured binding energies are two to three orders of magnitude smaller than that of the helium dimer, which is often considered the weakest bound diatomic molecule.

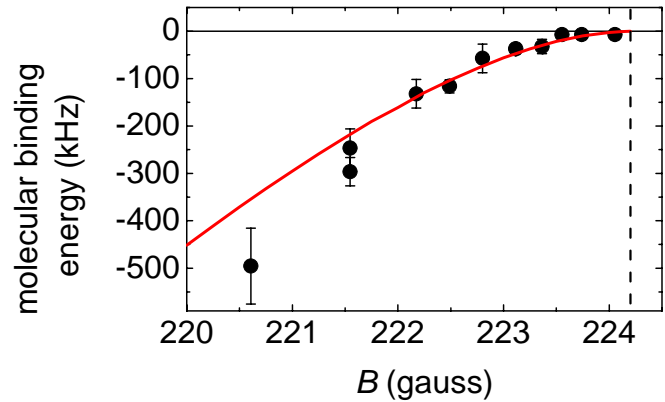

**FIGURE 4:** Molecules at a Feshbach resonance are extremely weakly bound objects with a binding energy in the kHz regime. Here, the binding energy, measured through rf photodissociation, is plotted versus the *B*-field. Despite the weak binding, these $^{40}$K dimers are amazingly stable objects at ultralow temperatures [10], surviving many collisions in an strongly interacting regime.

## THE BCS-BEC CROSSOVER

Let us return to the subject of making condensates in a Fermi gas of atoms. There are two seemingly very different pictures one could have of these condensates. The first is based on the well-understood phenomena of Bose-Einstein condensation of composite bosons. Here we imagine using the Feshbach resonance to convert the atoms pairwise into diatomic molecules, whose size is smaller than the average interparticle spacing. These molecules are bosons and so can form a Bose-Einstein condensate in much the same way as bosonic atoms. The second viewpoint arises from an analogy to low-temperature superconductivity, and relies on the well-tested "BCS" theory of Bardeen, Cooper, and Shrieffer. Here a weak attractive interaction results in the formation of Cooper pairs which simultaneously Bose condense. Cooper pairs are huge and cannot be thought of as independent particles since their size is typically orders of magnitude larger that the average spacing between fermions.

Much theoretical effort has gone into trying to understanding the connection between Fermi superfluidity (or superconductivity) and Bose-Einstein condensation. In 1980 a seminal paper by Tony Leggett proposed that these two pictures were limiting cases of a more general theory [18]. The essential difference between the two pictures is how tightly the fermions are bound into pairs. The first picture - a simple Bose-Einstein condensate of diatomic molecules – assumes that the pairs are so tightly bound that we can safely ignore the underlying fermions that make up our bosonic molecules. In contrast, the second picture assumes an extremely weak attraction between fermions, and the quantum behavior of fermions plays an essential role. In the so-called BCS-BEC crossover picture one expects that as a function of the interaction strength between fermions, condensation behavior in Fermi systems evolves smoothly between the well-understood limiting cases. Further, in this more general theory it is clear that both the fermion nature of individual particles and the boson nature of pairs must be considered on an equal footing (Fig. 5).

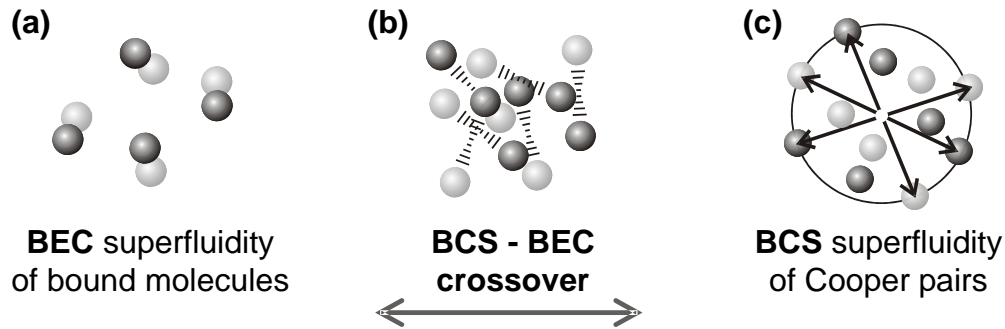

**FIGURE 5:** Superfluidity of Fermions requires pairing, since a pair of two Fermions has integer spin and can act as an effective Boson. Simply speaking, these pairs undergo Bose-Einstein condensation and form a superfluid. (a) The Bosons can be tightly bound pairs, for example two-body molecules. If the binding energy is much larger than the many-body energy scale, the fermionic degree of freedom does not play any role and the molecules simply form a Bose-Einstein condensate. (c) There can be a much more subtle kind of pairing, as observed in superconductors: BCS pairs are correlations of particles on the opposite side of the Fermi momentum sphere. The pairs, which are large in space, are an explicit many-body effect and only arise in the superfluid phase. (b) It has been proposed that the two seemingly distinct regimes of BEC of molecules and BCS superfluidity of Cooper pairs are continuously connected through a BCS-BEC crossover. The nature of the generalized Cooper pairs in the crossover is in between a molecule and a BCS Cooper pair: The pairs require many-body effects, but pairing is much stronger than in the perturbative BCS limit, giving rise to high-temperature superfluidity. The size of the pairs is on the order of the interparticle separation.

Now we see that the control over interactions afforded by Feshbach resonances in ultracold gases can be used to experimentally explore the predicted BCS-BEC crossover. With one system we can tune the interactions over what is arguably the most interesting regime of the crossover. This is the regime of strong attractive interactions where one would expect small Cooper pairs and also the regime of large molecules. That is, we can access the cusp of the crossover where the pair size is predicted to be comparable to the spacing between atoms in the gas.

The BCS-BEC crossover theory is potentially very powerful in that it would connect phenomena occurring over a tremendous range of pair sizes and transition temperatures. The range of interactions strengths accessible for an atomic Fermi gas near a Feshbach resonance is particularly interesting because it is predicted that high transitions temperatures will be reached near the cusp of the BCS-BEC crossover. Note that for the purpose of comparing systems with very different densities and masses, one should consider not the absolute temperature but rather the temperature normalized by the Fermi temperature.

# MAKING CONDENSATES WITH FERMIONS

It is the unique ability to tune interactions using a Feshbach resonance that has made it possible to create condensates with a Fermi gas of atoms. This amazing new tool provides the first experimental access to the region of the predicted BCS-BEC crossover which is neither described by BCS nor by BEC physics and where the condensation phenomena is not at all well understood theoretically. Moreover, the dilute atomic gas system turns out to be an exquisite, model system where it is hoped that theoretical understanding could be built up from first principles.

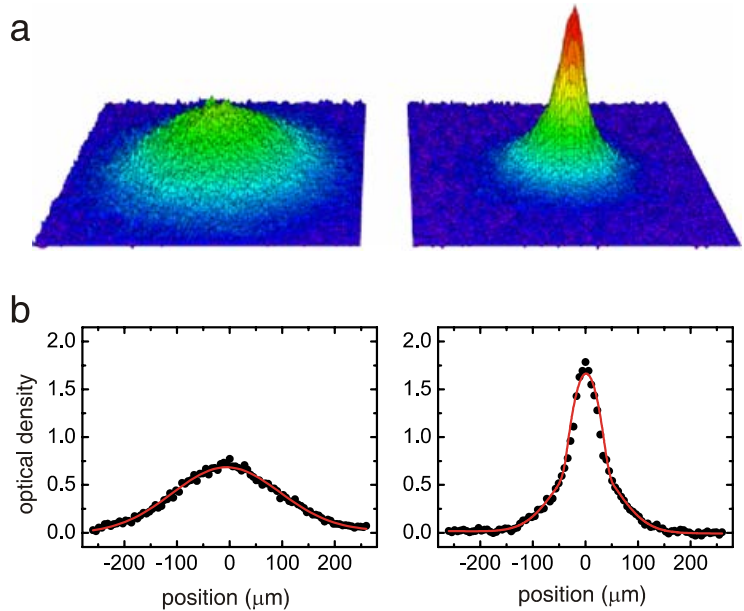

**FIGURE 6:** First time-of-flight absorption image of a molecular Bose-Einstein condensate: (a) Surface plots of thermal cloud (left) and BEC (right) with characteristic peak around zero momentum; (b) The thermal molecule cloud fits perfectly to a Gaussian. The molecular BEC shows the characteristic bimodal momentum distribution of a condensate plus a thermal cloud.

In the experiments we take full advantage of our unique ability to continuously tune the interaction strength in a single, clean system. In particular, we can, and do, change the interaction strength in the ultracold gas in real time. For example, let us consider two different approaches to creating a Bose-Einstein condensate of molecules, which were in fact simultaneously and independently realized in recent experiments. First, one could imagine a straightforward approach where molecules are created and then evaporatively cooling to a temperature below the critical temperature for the BEC phase transition. This was been done using $^6$Li and applying a magnetic-field strength very near the Feshbach resonance value but on the molecule side [13]. As the gas was evaporatively cooled, molecules were produced through collisional processes and then upon further cooling the molecule gas formed a BEC. Second, one could imagine a very different approach that probes the predicted BCS-BEC crossover physics and uses the fact that the interaction strength can be changed in real time. Here one starts with an ultracold Fermi gas of atoms and simply tunes the interactions to be more and more attractive until molecules are formed. Indeed, in our experiment with $^{40}$K atoms, we produced a molecule condensate with just this approach [12]. There was no direct cooling of the molecules, but instead we slowly changed the interactions strength in the original Fermi gas of atoms until a BEC formed on the molecule side of the Feshbach resonance.

It should be emphasized that in both the $^6$Li and $^{40}$K experiments, the "molecular BEC" is not a typical BEC. From the point of view of the BCS-BEC crossover these molecule condensates can already be far from the limiting case of a simple BEC of composite bosons. For the BEC theory to be valid, the bosons (the molecules) should be much smaller than the spacing between nearest neighbors in the gas. However, the Feshbach resonance molecules can be extremely large and for magnetic-fields very close to the resonance neighboring molecules have an increasingly large probability to overlap (as Cooper pairs do).

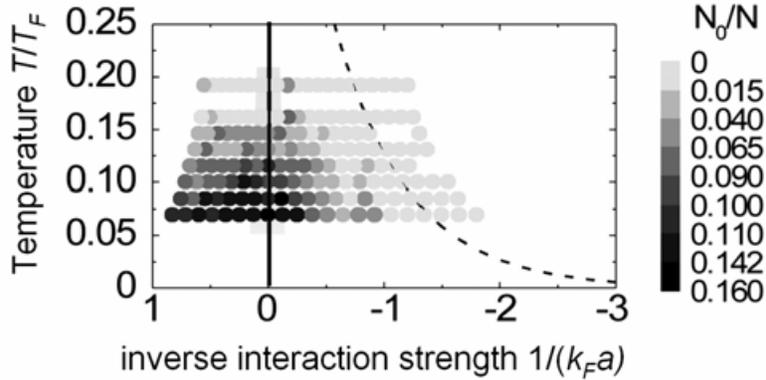

**FIGURE 7:** This phase diagram illustrates, through the universal parameter $1/k_F a$, the regime of the BCS-BEC crossover where our condensates occur. The measured condensate fraction $N_0/N$, indicated by the shades of gray, is plotted for different values of the initial $T/T_F$ and $1/k_F a$. Here $k_F$ is the Fermi momentum, $T_F$ is the Fermi temperatures, and $a$ is the interatomic scattering length. The temperature $T$ and $k_F$ are measured in a weakly interacting regime, from which we do a nearly iso-entropic $B$-field sweep into the strongly interacting regime. The dashed line indicates a calculation of $T_c/T_F$ for a homogeneous system in the BCS limit. Although the BCS theory that is not valid in the strongly interacting regime, $1/k_F a \ll 1$, this line provides an elucidating reference [19].

Fermi gas experiments are now actively exploring the region of the BCS-BEC crossover accessible via Feshbach resonance tuning of the interaction strength. In particular, we have shown that condensates exist on the atom side of the Feshbach resonance where there are very strong attractive interactions between atoms but no bound molecule state [1]. To be more precise, on this side of the Feshbach resonance (higher magnetic field for our $^{40}$K resonance) two atoms by themselves cannot form a bound molecule. Nevertheless in our Fermi gas we do observe condensates of atom pairs whose very existence then must depend on the fermionic nature of the atom gas. These pairs can be regarded as generalized Cooper pairs in the BCS-BEC crossover regime.

The experimental system with ultracold fermionic atoms gives the unique opportunity to experimentally study condensed matter questions, like the BCS-BEC crossover. At this time, the first studies of excitation spectra in this strongly interacting regime have been carried out [20-24]. Currently, experiments are trying to probe superfluidity by creating vortices and get a more quantitative understanding of the physics in this regime. We hope that future investigations, both experimental and theoretical, will help better elucidate the connection between Bose-Einstein condensation, superconductivity, and superfluidity.

## ACKNOWLEDGMENTS

We would like to thank J. T. Stewart for experimental assistance. This work was supported by NSF and NASA. C. A. R. acknowledges support from the Hertz Foundation.

## REFERENCES

1. Regal, C. A., Greiner, M., and Jin, D. S., *Physical Review Letters* **92**, 040403 (2004).
2. Anderson, M. H., Ensher, J. R., Matthews, M. R., Wieman, C. E., and Cornell, E. A., *Science* **269**, 198 (1995).
3. Davis, K. B., et al., *Physical Review Letters* **75**, 3969 (1995).
4. Greiner, M., Mandel, O., Esslinger, T., Hansch, T. W., and Bloch, I., *Nature* **415**, 39 (2002).
5. DeMarco, B. and Jin, D. S., *Science* **285**, 1703 (1999).
6. Loftus, T., Regal, C. A., Ticknor, C., Bohn, J. L., and S., J. D., *Physical Review Letters* **88**, 173201 (2002).
7. Regal, C. A. and Jin, D. S., *Physical Review Letters* **90**, 230404 (2003).
8. Regal, C. A., Ticknor, C., Bohn, J. L., and Jin, D. S., *Physical Review Letters* **90**, 053201 (2003).
9. Regal, C. A., Ticknor, C., Bohn, J. L., and Jin, D. S., *Nature* **424**, 47 (2003).
10. Regal, C. A., Greiner, M., and Jin, D. S., *Physical Review Letters* **92**, 083201 (2004).
11. Greiner, M., Regal, C. A., Ticknor, C., Bohn, J. L., and Jin, D. S., *Physical Review Letters* **92**, 150405 (2004).
12. Greiner, M., Regal, C. A., and Jin, D. S., *Nature* **426**, 537 (2003).
13. Jochim, S., et al., *Science* **302** (2003).
14. Zwierlein, M. W., et al., *Physical Review Letters* **91** (2003).
15. Zwierlein, M. W., et al., *Physical Review Letters* **92**, 120403 (2004).
16. Holland, M., Kokkelmans, S., Chiofalo, M. L., and Walser, R., *Physical Review Letters* **87**, 120406 (2001).
17. Donley, E. A., Claussen, N. R., Thompson, S. T., and Wieman, C. E., *Nature* **417**, 529 (2002).
18. Leggett, A. J., *J. Phys. (Paris), Colloq.* **41**, 7 (1980).
19. Note that the temperature values of the data is not the actual temperature in the strongly interacting regime, but measured in the weakly interacting regime. Therefore, the comparison of the BCS-temperature to the data is only qualitative.
20. Bartenstein, M., et al., *Physical Review Letters* **92**, 203201 (2004).
21. Chin, C., et al., *Science* **22**, 1100818 (2004).
22. Thomas, J. E., et al., *Journal of Low Temperature Physics* **134**, 655 (2004).
23. Kinast, J., Hemmer, S. L., Gehm, M. E., Turlapov, A., and Thomas, J. E., *Physical Review Letters* **92**, 150402 (2004).
24. Greiner, M., Regal, C. A., and Jin, D. S., *cond-mat/0407381* (2004).